# Socialbots and the Challenges of Cyberspace Awareness

Shashank Yadav[1]

**Abstract**

As security communities brace for the emerging social automation based threats, we examine the mechanisms of developing situation awareness in cyberspace and the governance issues that socialbots bring into this existing paradigm of cyber situation awareness. We point out that an organisation's situation awareness in cyberspace is a phenomena fundamentally distinct from the original conception of situation awareness, requiring continuous data exchange and knowledge management where the standard implementation mechanisms require significant policy attention in light of threats like malicious social automation. We conceptualise Cyberspace Awareness as a socio-technical phenomena with Syntactic, Semantic, and Operatic dimensions – each subject to a number of stressors which are exacerbated under social automation based threats. The paper contributes to the ideas of situational awareness in cyberspace, and characterises the challenges therein around tackling the increasingly social and often pervasive, automation in cyber threat environments.

**INTRODUCTION**

Social automation is generally understood as the use of software tools and techniques to automate and optimise the social interactions of a computational agent (Marinaccio et al., 2015; Nitta et al., 1993; Woolley, 2018). Not only it is ubiquitous in cyberspace, the highly standardised nature of online platforms (uniform profiles, quantifiable and predetermined range of actions) leaves little for humans to distinguish themselves from bots that access all the same environmental attributes of cyberspace (Bakardjieva, 2015). Moreover, in their natural habitat, these bots have what most huamans lack – the scale, the speed, and a tireless goal-driven existence. Owing to this the scholars and practitioners of cyber security and propaganda alike had been long warning about the now prevalent automation of social engineering (Ariu et al., 2017; Duff, 2005; Frumento et al., 2016). Following from a unified conception of social engineering as the psychological manipulation of people to get them to exhibit certain thoughts and behaviour, we define Malicious Social Automation (MSA) as the use of social automation to exploit or enable the exploitation of human psychology in order to achieve the objectives of a threat actor. The agents of this are socialbots, automated accounts which imitate online human behaviour (Boshmaf et al., 2013).

---
[1]  shashank@asatae.foundation

The applications of MSA can be further understood via three overarching operational constructs – the primary Memetic and Transgressive types – where Memetic automation encapsulates the social automation aiming at the desired spread of messages and ideas, and Transgressive type encapsulates the social automation that violates the established security assumptions of an environment/user. And a third, Supportive type of MSA primarily aiming at facilitating and coordinating the Memetic and/or Transgressive operations, such as user profiling or Command and Control (C2) management over social platforms.

Since 2021, Microsoft's threat intelligence reports also began to consider online psychological operations as disruptive to the enterprise and nation-state cyber security (Microsoft, 2021). Much of these online psychological operations consist of 5% to 10% socialbots that tie together various small networks and alter the natural assortment of influence (Cheng et al., 2020; Stewart et al., 2019; Zhang et al., 2022). In fact, many applications of MSA defy conventional cyber security thinking since most of user interactions with online information flows and social agents may occur outside an organisation's own digital network – which presents additional challenges to maintaining cyber situation awareness. As the cyber threat environment itself is dominated by Advanced Persistent Threats and automated attacks, including from sophisticated first-world military organisations (Graphika & Stanford Internet Observatory, 2022; Meta, 2022), it is pertinent for defending governments to maintain persistent and actionable situation awareness in cyberspace. This has led to an increased requirement worldwide for automated mechanisms to maintain continuous cyberspace awareness among various stakeholders, exemplified by the operationalisation in 2021 of a structured-information based bidirectional automated Threat Intelligence eXchange platform by the Indian Computer Emergency Response Team (CERT-In) (CERT-In, 2020). In 2022, North Atlantic Treaty Organisation (NATO) began to back its own language and information standards for specifying foreign Information Manipulation and Influence (FIMI) operations, which are also based upon the same standardised information models and exchange mechanisms (StratComCoE, 2022).

Further, automated information exchanges have long been desired by government structures like Computer Emergency Response Teams (CERTs) and various Security Operations Centers (SOCs) for communicating and coordinating on various cyber situations (Dandurand & Serrano, 2013), and having to maintain a common operating picture and desired readiness levels. Internationally accepted information standards enable and make this endeavor a lot seamless. However, the dependence over stakeholder communication and automated Cyber Threat Intelligence (CTI) sharing mechanisms also puts cyber situation awareness in a different league than the conventional ideas of situation awareness, with broad implications for national policies and cyberspace governance when it comes to MSA type threats.

**THE NATURE OF CYBERSPACE AWARENESS**

Knowing what is going on around you, or Situation Awareness, is a psychological construct originating from air combat (Endsley, 1995). Deconstructed into the stages of perception, comprehension, and prediction – the original conception of situation awareness is largely about various stages of mental representation of the environment, suited to analyse the decision making of the human operators of a system. The model only works well where a technology operator can perceive, plan, and act upon an environment, and the feedback of that action is available to optimise the operator's world-model for further planning and action, similar to the workings of any reinforcement learning agent.

Owing to this limitation, theoretical development of the concept have taken three routes to conceptualisation – a) the original, where situation awareness is in the mind, b) the technocratic, where situation awareness is a knowledge artifact (Mulgund & Landsman, 2007), and c) the integrative, where situation awareness is an emergent phenomena of interactions in a wider socio-technical system (Stanton, 2016). A practitioner of the second approach, when asked where is the situation awareness, might say that it is "right on the screen". This is consistent with the modern practices in cyber security where much of the efforts towards creating situation awareness lead into visualising the incoming situational data (Franke & Brynielsson, 2014). Notwithstanding, we are going to conceptualise situation awareness in cyberspace, or 'cyberspace awareness' for brevity, through the last approach which is also known as a distributed situation awareness approach. It is fitting because no single organisation has all the information needed to form a definitive real-time picture of cyberspace, making communication and collaboration essential to knowing.

Moreover, cyberspace itself, as patently experienced, isn't an environment which humans are evolutionary designed to perceive–comprehend–predict about. Other than visual-auditory predetermined alarm signals (Endsley et al., 2003) or intuition resulting from years of experience, an analyst who is inundated with millions of threat indicators everyday has little recourse for judging the actual situation of cyberspace. As CTI forms the basis of any actionable cyberspace awareness and this situational information needs to be shared with a wide range of stakeholders to develop a common operating picture, market forces and governments try to address this problem with curated CTI feeds (Bouwman et al., 2020). Since CERT-In's CTI feeds are based upon a standardised, open-sourced, and extensible language called Structured Threat Information eXpression (STIX) (CERT-In, 2020), we will undertake this as the standard mechanism for building automation-centric cyberspace awareness while exploring its intersection with the threat of MSA.

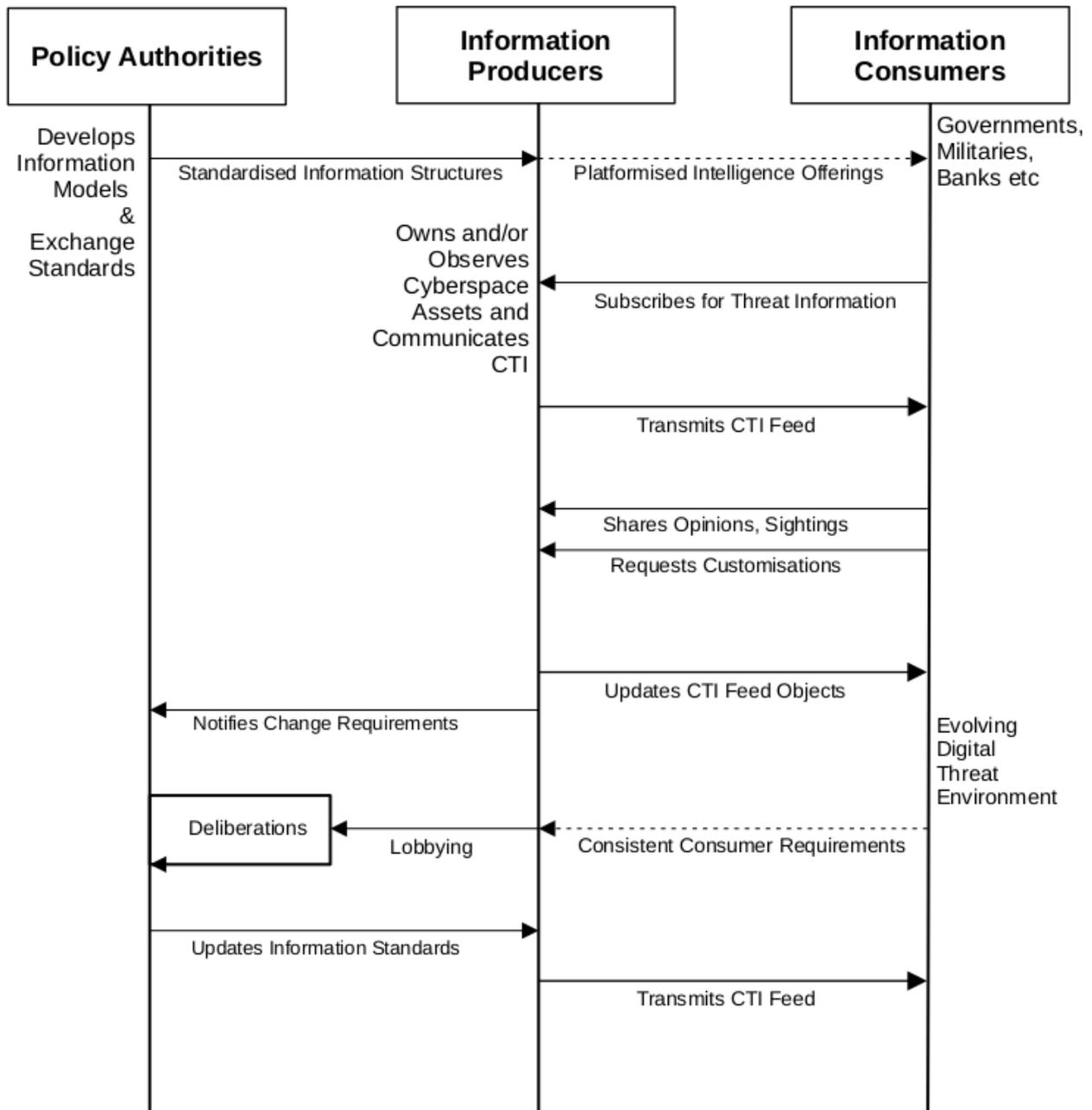

*Figure 1: A simplified interaction between CTI stakeholders*

STIX is designed to be machine-readable so that these feeds can be directly plugged into an organisation's automated response or alerting systems. For actual sharing of thus appropriately formatted information, stakeholders have to rely upon Trusted Automated eXchange of Indicator Information (TAXII), which is a transport standard that defines how that information is shared (Jordan et al., 2021). In a nutshell, STIX defines the content of a cyber situation and TAXII defines its transmission. Both of these standards are presently managed by the Organisation for Advancement of Structured Information Standards (OASIS) which acts as a policy authority along with other standard and industry bodies. The entire ecosystem also draws significantly from knowledge bases created and maintained by MITRE, a security non-profit that had started as a contractor for Defense Advanced Research Projects Agency (DARPA) and later grew into a leading cyber security evangelist (Day, 1996).

The functioning, the constituency, and the strategic interests of these organisations are not the only concerns while undertaking a critical analysis of how good the present automated mechanisms of building national cyberspace awareness are to combat malicious socialbots. The organisational knowledge management practices around acquiring and processing actionable CTI itself consists of three dimensions – Syntactic, Semantic, and Operatic – examining which sheds the much needed light on the nature of cyberspace awareness. It also draws out the specific issues that emerge around malicious socialbots in data policy environments, cognitive environments, and regulatory/strategic policy environments respectively.

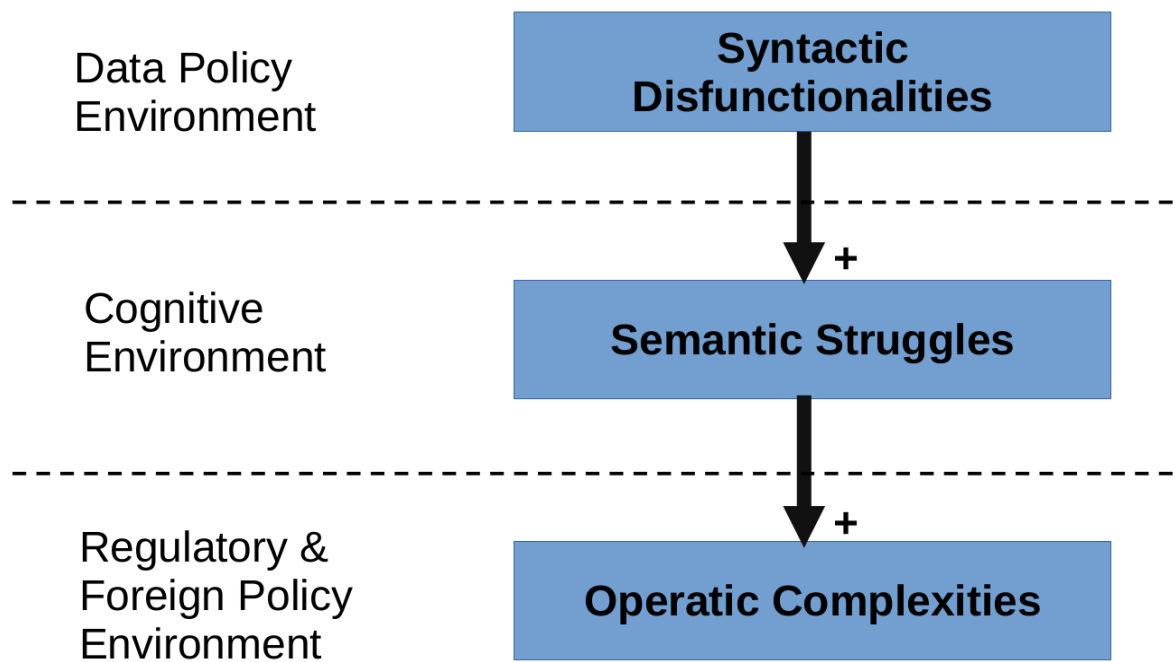

*Figure 2: With growing threats of social automation, states need to be mindful of all the three dimensions of Cyberspace Awareness*

It is noteworthy that while situation awareness itself has never been conceptualised in this manner yet, this three-dimensional approach to knowledge itself isn't completely new (Carlile, 2002). Information and communications have been traditionally viewed through 'semantic' and 'syntactic' lenses. While Shannon had considered the semantic aspects of information as irrelevant to the engineering problem (Shannon, 1948), those did matter considerably within organisational theory (Nonaka, 1994), which prompted some scholars to further discuss 'knowledge in practice' (Carlile, 2002). We will integrate all the three dimensions of knowledge into a cohesive formulation of cyberspace awareness, and thereby not only viewing digital threat information through the commonly accepted syntactical and semantical paradigms, but also introducing an Operatic dimension of it, which pertains to its orchestration towards meeting the organisational or national goals and interests. We'll now elaborate the challenges presented by MSA to states' cyberspace awareness through each of these three dimensions.

**THE SYNTACTIC DIMENSION OF CYBERSPACE AWARENESS**

The syntactic dimension of cyberspace awareness pertains to what aspects of digital reality are chosen to describe and/or communicate it and how they are encoded. That was the main concern of the discussions on IDXWG email list, established by CERT-United States and CERT.org to streamline automated data exchanges regarding cyber incidents, from which STIX grew out (Barnum, 2012). Today as most CERTs have embraced the mechanism, it is important to note that this also reflects the increasing dependence on automated systems for handling cyberspace situations. Even in case of purely memetic situations, the growing use of automated content moderation practices (Gillespie, 2020) points at the instrumentality and necessity of examining the underlying information models that generally enable these alerting or policy enforcing software tools to communicate with each other.

The underlying information model (or the syntax of cyberspace awareness) for STIX has gone through multiple changes, but while its open-source, changes to the model require a significant degree of lobbying[2] and networking. Any proposed changes to the existing syntax have to go through, and made by, the OASIS Technical Committee (TC). For these changes to be integrated into the standard information model, a 'call to consent' (or 'to object') is issued. The consent of committee members is assumed unless one registers an objection. At which stage the objecting member has to a) indicate the objection on a ballot, and b) provide a reason for the objection and its proposed remedy to the TC. The whole process is executed via an online voting system, and there is also an invitation to comment before the call for consent is issued. The publicly available OASIS mailing list for CTI users suggests little public discussions within the community over these decisions.[3] Ironically, much of the interim processes of developing the open-sourced cyberspace awareness mechanisms are kept closed to the public eye due to intellectual property concerns[4].

Here, it is important to assess how good the existing model is, or how much customisation it may require, to represent MSA based threats. We'll borrow from Situation Theory that approaches understanding a situation as a context modeling excersice by combining discrete informational items under a relation (Devlin, 2006). Such a characterisation of 'the situation' as a syntactically formalised composite information structure, suitable for human as well as machine reading also validates prevalent models of analysing cyber security events. Thus , fitting into the widely used

---

[2] Some change requirements can be seen at https://github.com/oasis-tcs/cti-stix2/issues
[3] https://lists.oasis-open.org/archives/cti-users/ ; https://lists.oasis-open.org/archives/cti/
[4] Discussions on OASIS CTI TC public email thread with subject "Inviting nominations for Chair of Cyber Threat Intelligence (CTI) TC"

Diamond Model of Intrusion Analysis, we translated MSA into a formal schema that can be populated with data from CTI feeds, and assessed the issues of its orchestration into the present standards-based structured information feeds.

Interestingly, the 'diamond model' of analysis was first brought forward by Michael Porter, an economist, to show relations between the factors that lead to national competitive advantages (Porter, 2011). It was eventually repurposed by American Department of Defense into the "diamond model of intrusion analysis" (Caltagirone et al., 2013). In the diamond model of MSA, we have four entities: Threat Actor, Target, Infrastructure, and MSA; connected by a 'campaign' relation.

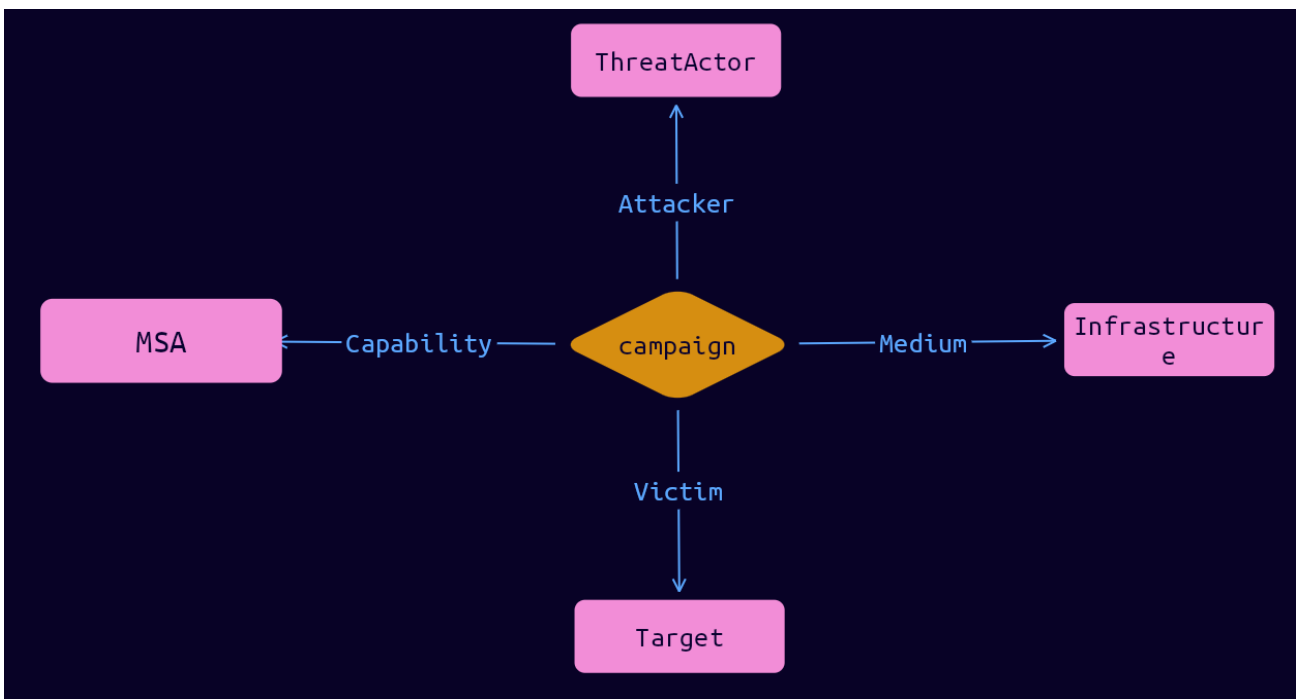

*Figure 3: Related by being in an MSA campaign relationship, the Threat Actor plays the role of an Attacker, the Target plays the role of Victim, Infrastructure plays the role of Medium, and MSA plays the role of Capability in the 'diamond model' of MSA. [Expanded image in Appendix A]*

| Object<br>*[Parent]*<br>*[Parent:Child]* | Attributes | Description | Note |
|---|---|---|---|
| Threat Actor | ID | Auto-generated String | |
| | Name | Name is assigned, and can differ from one producer to another | |
| | Capacity | Same as STIX Resource_Level | |
| | YoE | How old or new a threat actor is | |
| | Objective | Strategic objective behind actor's activities | |
| | MITRE_ID | Mapping the actor in MITRE database | |
| Threat Actor: APT | Affiliation | Nation/State affiliation | |
| | Support_Type | Level of state support | |
| | Human_Resources | Offensive organisation/trollfarm size | |
| Infrastructure | ID | Auto-generated string | |
| | Infrastructure_Type | Type of infrastructure, same property as in STIX Infrastructure SDO | |
| Infrastructure: Logical | Mainbots | Name/IDs of key automated accounts | |
| | C2 | Command and control method | |
| | Platform | Name of online platform | |
| | Toolkit | Software stack i.e. OS, exploit kit etc | |
| Infrastructure: Physical | Device | Hardware description | |
| | Location | As implied, same as STIX Location SDO | |
| MSA | ID | Auto-generated String | |
| | C2 | Shared property with logical infrastructure | |
| | Intent | Tactical objective for using MSA | **This set of attributes could form an independent domain object describing automated influence operations.** |
| | Supportive_AI | Possible type of ML support, i.e. image, language etc | |
| | bot_actions | Structural capabilities i.e. post, share, like etc | |
| | bot_numbers | Estimated number of bots | |
| | bot_roles | Generator, Short Term Supporter, Long Term Supporter | |
| MSA: Memetic | Start_Date | When the campaign began | |
| | End_date | When the campaign ended/ongoing | |
| | Platform | Shared attribute with logical infrastructure | |
| | Topic | Topic of discourse | |
| | Keywords | Keywords identifying the narrative | |
| | MCF | Main Creative Format | |
| | SCF | Secondary Creative Format | |
| | Landing_Page | Where most traffic is directed at | |
| | Engagement_Level | Requires standardised dictionary to reflect public engagement i.e. Viral, Organic etc | |
| | Impressions | Estimated reach of messaging at the time of CTI creation | |
| | Impressions_Pro | Projected reach of messaging | |

| MSA: Transgressive | First_Observation | When was the transgressive activity first observed | |
| --- | --- | --- | --- |
| | Intrusion_Type | Type of security violation | |
| | Malware_Type | Similar to Malware_Type in STIX | Needs updated vocabulary w.r.t. bot definition and types |
| | IOC | Observables, content identification | Tools like Meta ThreatExchange's open source Hasher-Matcher-Actioner can be used for content id |
| | CKC_stage | Stage of attack in Lockheed's kill chain | |
| Target | id | Auto-generated String | |
| | Current | Present target | |
| | Lateral | Target the attack could expand/jump to | |
| *Capitalisation in table for increased readability. Excludes common autogenerated properties, i.e. creation time etc | | | |

The current STIX has 18 core STIX Domain Objects (SDO) (Jordan et al., 2021). Each SDO represents a standalone concept commonly used in cyber security, i.e. threat actor, vulnerability, tool etcetera. A lot of the time, a CTI producer sharing information about a new or out-of-context threat may have to add custom properties to core SDOs, or sometimes even create custom SDOs if required. While the language itself is fairly flexible and extensible so that new ad-hoc objects can be created, increased use of non-standard custom objects creates conditions for standardising their syntax, to reduce uncertainty and semantic conflicts.

For example, in the case of MSA's information model described above, we can argue that memetics could be a standalone SDO, which would imply some policy dilemmas in its orchestration. Afterall most malicious campaigns, whether informational/psyop or not, do involve communications which could be described in a formalised syntax like automated advertising is described to the ad server by a tracking pixel. These become more relevant when the malicious campaign involves large scale disinformation and social engineering tactics. Therefore only to emphasise the characteristics of communications under an independent 'memetic' MSA object, we kept properties such as bot_roles, bot_numbers, and bot_actions under the parent type of MSA.

Hypothetically, if a STIX extension describing the memetics of a malicious campaign becomes widespread, it will be competing with other similarly named and designed information objects. For example, if a customised Memetic SDO was to be used by a government CERT to transmit information about the contents of MSA communications, it is very much possible that other government organisations and private CTI producers may also start transmitting similar SDOs, albeit with conflicting names/meaning/specifications. Thus a need for standardisation of such an extension, or its inclusion into STIX core objects would quickly emerge.

This unrestricted syntactical extensibility of cyberspace awareness languages, when widely adopted to describe the same realities, breeds grounds to curb the effects of that extensibility by revising existing standards. However, the change management concerns around the syntax of cyberspace awareness are irredeemably intertwined with its intellectual property concerns. While the STIX/TAXII standards are managed by OASIS, the Intellectual Property Rights (IPR) remain with US Department of Homeland Security. And based on TC's discussions (OASIS, 2022), everytime the standard is revised, the IPR needs to be waived so that OASIS can get a special license to operate these. This issue has come up frequently in the OASIS' cross standardisation efforts with the International Telecommunication Union (ITU) (OASIS, 2022).

**THE SEMANTIC DIMENSION OF CYBERSPACE AWARENESS**

Writing in Data and Reality, William Kent remarks that form and content (or syntax and semantics) are mixed, they affect each other, but the semantic significance is only truly perceived by the end user and not the producer of data (Kent, 2012). The end user here, an analyst who has to quickly escalate a relevant threat up in the organisational chain, often finds the incoming data to be incomplete, unreliable, and even subjective (Yucel et al., 2020). The semantic dimension of cyberspace awareness pertains to the sense-making and interpretation of this "high velocity – high volume" cyber situational data.

The consumers of CTI feeds, who are expected to maintain the requisite cyberspace awareness for their organisation, have to face primarily three kinds of semantic gaps. First where the incoming data objects fail to create a useful picture of the situation, second where the vocabulary used within the feed creates ambiguity and conflicts over the usage and implications of a term, and third where the feed instances themselves are poorly labelled.

- *Meaningfulness of Situation Transmissions*

Of the various types of STIX objects, a CTI producer can choose to send any number of them, while also adding relationship objects to signify how one object is connected to the other. However such bundling of objects is left at the discretion of the information producer, and is not mandatory to implement (Jordan et al., 2021). This is fine for atomised information of a malware or hardware vulnerability but the threats of social automation are more contextual and their experience can vary from one target to another, therefore meaning-making is integral to the process of risk assessment of MSA. Further, the information sharing and language standardisation itself is silent on the structure of a situation, i.e. to make an analogy with natural languages, its grammar says little on 'sentence formation' and is mostly focused on 'word formation'.

Unlike an atomised CTI object, a cyber 'situation' on the other hand is a function of specific and discrete objects that describe and contextualise a specific event in time. Lack of any one of those informational objects hinders the development of the big picture. For example, following the diamond model, we can say that,

Situation = (((Threat Actor) x Probability$_{\text{Threat Actor}}$)), ((Infrastructure) x Probability$_{\text{Infrastructure}}$)), ((MSA) x Probability$_{\text{Campaign}}$)), ((Target) x Probability$_{\text{Target}}$)))

Above can be read in simple English as a Threat Actor using some Infrastructure to deploy MSA against a Target. There is also a confidence score associated with every object, which is a common STIX property. But if any of the objects making up the situation of MSA, or their accurate subtypes or relevant properties are missing, the new information would only add to the existing ambiguity for the organisation, and perhaps incur some technical debt in resolving that ambiguity as well. We see that the present information sharing standards, while aiming for situation awareness, are better suited for atomised data sharing, and the pursuit of situation awareness is left to the analyst and his organisation, consistent with the traditional conception of situation awareness existing in the mind (Stanton et al., 2010).

This lack of rules or integrity constraints that enforce actionable meaningfulness has led to some commercial CTI producers compensating for the lack of information quality with more quantity of data, effectively offering feed updates at very short intervals. However, scholars note that the burgeoning number of CTI vendors and larger intelligence feeds do not translate into better data (Li et al., 2019), and consequently meaningful cyberspace awareness remains largely elusive.

- ***Contested Open Vocabularies***

OASIS has created several open vocabularies under STIX. It should be noted that in the earlier version of STIX, these were called 'Default Controlled Vocabularies' (Barnum et al., 2016) and the use of the word 'Open' for the same, although there is negligible operational difference between them, only began with the second version of STIX when the first version was merged with the MITRE's proprietary Cyber Observable eXpression (CybOX) to create a new dominant standard for sharing cyberspace awareness. These provide acceptable terminology for domain objects' properties and define the meaning of those terms.

However, the conventional cyber offensive tools/techniques and MSA have a slightly differing operational construct – while the former targets data, computers, and digital communications, the latter targets the minds of the communications' receiver. Consequently, many of the terms used while describing the former can have completely different meaning when used in context of the latter. A good example is the word 'Amplification' – In context of FIMI operations, the word 'amplification' is generally understood as bots or trolls catalysing a specific narrative, and is often used in discourses about disinformation (Peck, 2020; Phillips, 2018). However, in STIX open-vocab and traditional cyber security discourse, 'amplification' is often used to refer to DDoS kind of "flooding attacks" that render something inoperable (Anagnostopoulos et al., 2013; Ryba et al., 2015). So an analyst receiving information about "Amplification Infrastructure" has to know what type of amplification is being discussed, or if both meanings of amplification are relevant to the cyber situation at hand. Consequently, requisite changes need to be made in the standardised open vocabulary to make it desirably expressive in such threat situations and reduce the semantic ambiguity for analysts already overburdened with threat indicators.

- *Poor Labelling*

A lot of "high velocity" commercial CTI, including from prominent sources, comes with significant number of indicators just labelled as 'suspicious' or 'malicious' (Li et al., 2019). Ostensibly, it is up to the analyst to examine each item to ascertain whether any, and how much of, risk a particular cyber situation poses, and thereby often information consumers end up also subscribing to smaller but more pricey CTI which is better labelled and curated (Bouwman et al., 2020). Nevertheless, the larger practice of ambiguous generic labeling has been a major component leading to significant technical debt in CTI pipelines, as with more threat indicating information flooding an analyst, there is a significant proportion of false positives as well (Bourgue et al., 2013).

While the existing challenges of labelling are far from being resolved, MSA presents much more subjective and contested labelling environment. NATO's STIX-based DISARM framework, which intends to be a standardised language to share and communicate information about foreign influence threats, contains myriad of examples for the same (DISARM, 2022) – many of the offensive Tactics, Techniques, and Procedures (TTPs) listed under DISARM are rather benign and waste of a cyberspace analyst's time, such as "T0059: Playing the Long Game", "T0131.001: Legacy Web Content", "T0114.002: Traditional Media", "T0132.003: View Focused", "T0023.001: Reframe Context" etcetera. At the same time, its defensive TTPs can be interpreted to be rather offensive, for example, "C00052: Infiltrate Platforms", "C00048: Name and Shame Influencers" among many others. It stands true to the wisdom that there is little difference between offence and defence in information operations, however, the language and labelling used for many of the offensive TTPs gives an impression of anything but a benign everyday exchange on the internet.

If CTI about social automation based threats is to become common it requires much greater standardisation efforts for semantic clarity than presently have been mustered, as well as international efforts to create consensus on labelling based on objective rational examination of human affairs and the technical nature of information, and certainly not political allied/axis bloc-formation efforts, which brings us to the operatic dimension of cyberspace awareness.

**THE OPERATIC DIMENSION OF CYBERSPACE AWARENESS**

The operatic dimension of cyberspace awareness concerns neither the formulation of information, nor its understanding, but its orchestration – sharing it, holding it, making use of it to pursue organisational or national security objectives. This is an often overlooked area of public policy, because to a large extent, the production of security in cyberspace has been a private sector endeavor (Carr, 2016; Hiller & Russell, 2013). This occurs in stark contrast to the physical world where the production of security, to a large extent, is monopolised by the state via its sovereign right to use legal violence. The cyberspace however, is an arena where much of the activity happens over privately owned infrastructure – whether it is the operating system, networking devices, social platforms, the cloud or even much of the financial activity and critical infrastructure etcetera – most of it is orchestrated over non-state infrastructure (Carr, 2016). Consequently, private actors also become first responders to any untoward event, which has led to a range of such actors also developing their own CTI platforms, as well as monitoring for and broadcasting their own CTI (Samtani et al., 2020). Unsurprisingly, the biggest consumer of this information, which forms the basis of cyberspace awareness, are governments and military organisations (Bouwman et al., 2020), who use this information to further enrich government organisations like CERTs own cyberspace awareness and threat estimates. Nonetheless, for knowing and governing cyberspace, a complex and informal institutional architecture has taken shape where governments and private sector, and even civil society, often follow different incentives, unwritten rules, and transnational influences to collectively carry out the production of security and intelligence in cyberspace.

With the emergence of non-traditional cyber threats like MSA, this quagmire of tacit understandings and territorial jurisprudence (Kuehn et al., 2020) requires a deeper investigation in light of how various nation-state stakeholders orchestrate their cyberspace awareness. Unfortunately, the cyber information sharing and classification structures among stakeholders as well as the resultant complexities of international law and politics, which were already a maladjusted area of the existing system of cyberspace governance (M. N. Schmitt & Watts, 2016), get further dysfunctional under the new conditions of malicious and pervasive social automation.

- *Dilution of National Diplomacy into Global Governance Structures*

OASIS TC member meetings are also a record of TC worrying over standards competition with China, and members state that one of the benefits of submitting to the ITU is that it is a UN agency which would help with global adoption and overcoming Chinese ambitions (OASIS, 2022). In terms of global standards competition, G7 has already declared its "opposition to any government-imposed approaches... to reshape the digital technical standards ecosystem" (G7, 2021) – which suits the American approach towards standardisation which is designed in a manner that the process largely remains in the hands of the industry and market forces (DeNardis, 2011). While conventional cyber security issues have been less political, social automation, particularly the memetic variety, could be inherently political, and state actors unable to use conventional diplomacy will try to establish 'an order' that suits their interests through non-formal lobbying into global governance structures. An example of which is the NATO's Strategic Communications devision evangelising the DISARM framwork through a non-profit[5].

This non-formal mechanism of technology governance is also reflected in the constituency of OASIS (see Appendix B), where a multistakeholder setup comprising largely of western private corporations (there is 1 Chinese company and Japan is treaty-bound with US interests), and a few representatives of the US security establishment and academia deliberate over the required changes to the standard. Clearly, a multilateral approach towards syntactical development could alleviate some of the international operatic concerns as these cyberspace awareness standards get considered for wider international standardisation under the ISO/ITU.

- *The Dilemmas Over Cyber Proxies*

It is widely acknowledged in discourses on cyber defence and security that the rules of engagement or conduct of hostilities between states, as laid down in the Geneva Conventions are not suited to deal with present realities (Jeutner, 2019; M. Schmitt, 2012). A similarly misplaced institution can is also the UN Charter (Benatar, 2009). These pre-internet artifacts of 'global order' are often invoked whenever a discussion of laws surrounding cyber conflicts or governance is undertaken. And rightly so, the only other significant international endeavor in cyberspace regulation is the Tallinn Manual, which describes itself as an "expression of the opinions" (M. N. Schmitt, 2017b), leaving no doubt about a pervasive institutional bankruptcy in cyberspace governance, with wide implications for threat information sharing structures.

A good vantage point to understand this in context of social CTI is the DISARM framework, as it lays down a STIX-based common language to share threat information about information manipulation and interference operations. Consequently, a CTI producer could use it to describe many applications of MSA also, such as automated psyops, astroturfing bots, AI-mediated spearphishing etcetera. However, a socialised malware is a different beast than a malware. Its threat is not experienced uniformly, i.e. a social botnet strategically posting (and generating) pro-palestine memes and "watering-hole" links maybe considered a threat to Israeli interests but its threat perception would be far from uniform and perhaps nill (or even negative) among the Arab stakeholders. A traditional botnet, on the other hand, is a threat regardless of whether it is running on an Israeli computer or an Arab one. Thus, a commercial CTI producer broadcasting worldwide the traditional botnet's specifications via STIX/TAXII incurred no subjectivities and conflicts of interests, but packaging a 'social botnet' into a global threat matrix is riddled with political undercurrents and arguably only leads to the continuation of a political-military conflict via other means. Such 'other means', in the pre-internet universe of Geneva conventions, could well have been considered under the moniker of diplomacy, but can now be seen as "paradiplomacy" (Melissen, 2016).

---

5   https://www.disarm.foundation/

What the leading mechanisms of structured threat information broadcasting offer in the above scenarios is an Opinion SDO which can be attached to a contentious CTI object and relayed to the rest of the network (Jordan et al., 2021). However, most countries lack a government organisation to contest such poorly or strategically disseminated CTI objects. For states to effectively address this issue, other than sharing "opinions", they'd have to begin from the syntactical stage of cyberspace awareness and specify how a threat such as MSA should not be expressed and also propose the specific protocols that need to be followed in transmitting such information. However, we've seen that the current processes of CTI standardisation are exclusionary to most of the state actors (G7, 2021).

Nevertheless, private actors releasing (or contesting) social CTI over a global network will continue forward the slow erosion of traditional diplomatic methods of engagement between nation-states (Hocking, 2004). It should be noted here that under particularly political situations, a government organisation broadcasting social CTI of an alleged application of MSA can also be considered as breaking the principle of non-intervention (Jamnejad & Wood, 2009). Therefore, non-state actors and private corporations also serve as useful proxies in what some scholars have referred to as legal grey zones (M. N. Schmitt, 2017a) while referring to political-military use of information in cyberspace. Notwithstanding, the erstwhile absolute sovereignty of states and non-intervention, which had been the very basis of international law, may stand to get further eroded with states and their proxies taking automated CTI-based countermeasures to combat active social botnets.

- ***Information Sharing and Classification Structures***

Unlike the situational awareness of air combat, building an organisation's cyberspace awareness is participatory activity. Various governmental CERTs keeping their threat exchange bidirectional is only the tip of an iceberg of the continuous exchange of information required to truly know what the situation of cyberspace is (Dandurand & Serrano, 2013). The nature of information being potentially sensitive and security-centric, the information classification and sharing structures become important for information consumers (Murdoch & Leaver, 2015). The standard mechanisms competing for adoption, as well as CTI vendors and platforms, present their own approaches for the same. For example, while STIX does not concern itself with classification,

STIX-compliant platforms like MISP (Malware Information Sharing Platform, a widely used open-sourced CTI sharing software, which is also a government backed initiative by CERT-Luxembourg) do enable different sharing scenarios and customisable reach (C. Wagner et al., 2016) as well as pseudo-anonymous and encrypted mechanisms[6] to share threat information. Leading domain-specific international organisations like Forum of Incident Response and Security Teams (FIRST) have further outlined an Information Exchange Policy defining various sharing restrictions and handling of information (i.e. encrypt-in-transit) (MacDonald et al., 2019).

However, none of this constitutes any legally enforceable framework at all, and is akin to the Tallinn manual in the way that it is an expression of opinions that CTI vendors may consider implementing, or may not. Needless to say, nation-states and their jurisprudence about information classification and restriction have not yet found any relevance here. However, as the use of social automation gets increasingly common, governments and private sector will have to find a common ground over handling and classifying the threat information about automated social activities while also creating appropriate institutional mechanisms for monitoring and evaluating such information. The change of operational construct to bots being maliciously social may require more emphasis on trust and attribution and less on anonymity as prevalent (Murdoch & Leaver, 2015).

Since different cyber security vendors may subscribe to different information models and restriction mechanisms, and the lack of a unified legally enforceable information classification and restriction mechanism while sharing social CTI is bound to create 'human concerns', even if away from the eyes of the masses. A socialbot interacts with and influences real human users online, thus a comprehensive detailing of its activity may contain information, direct or inferred, that may be considered a breach of privacy and users' trust. So it is worth noting that there isn't any internationally enforceable information classification and encryption regime when sharing cyber threat information, moreover, various national-level data protection regimes can have incompatibilities that make legal CTI released in one country, completely illegal in another (T. D. Wagner et al., 2019). This is bound to be exacerbated with MSA, as sharing social CTI could also be considered as shaping political preferences (Huckfeldt & Sprague, 1987). Moreover, this also points at the bigger problem of the non-applicability of national secrecy regimes over the flourishing commercial and open-sourced transnational intelligence collection practices (The Economist, 2021).

**CONCLUDING REMARKS**

Humans are not evolutionarily equipped to have awareness of what is going on in the world of zeros and ones, let alone process millions of abstruse threat indicators streaming everyday. Thus, unlike the conventional individual centric conception of situation awareness, building cyberspace awareness for the modern state requires creating and maintaining persistent communication channels with a variety of stakeholders and commercial computer intelligence providers scattered across the globe. In effect, maintaining cyberspace awareness requires states to undertake a continuous management of knowledge as well as the knowing. Consequently, we see syntactic, semantic, and operatic issues emerging in the production and management of situational knowledge.

---

6   https://www.misp-project.org/

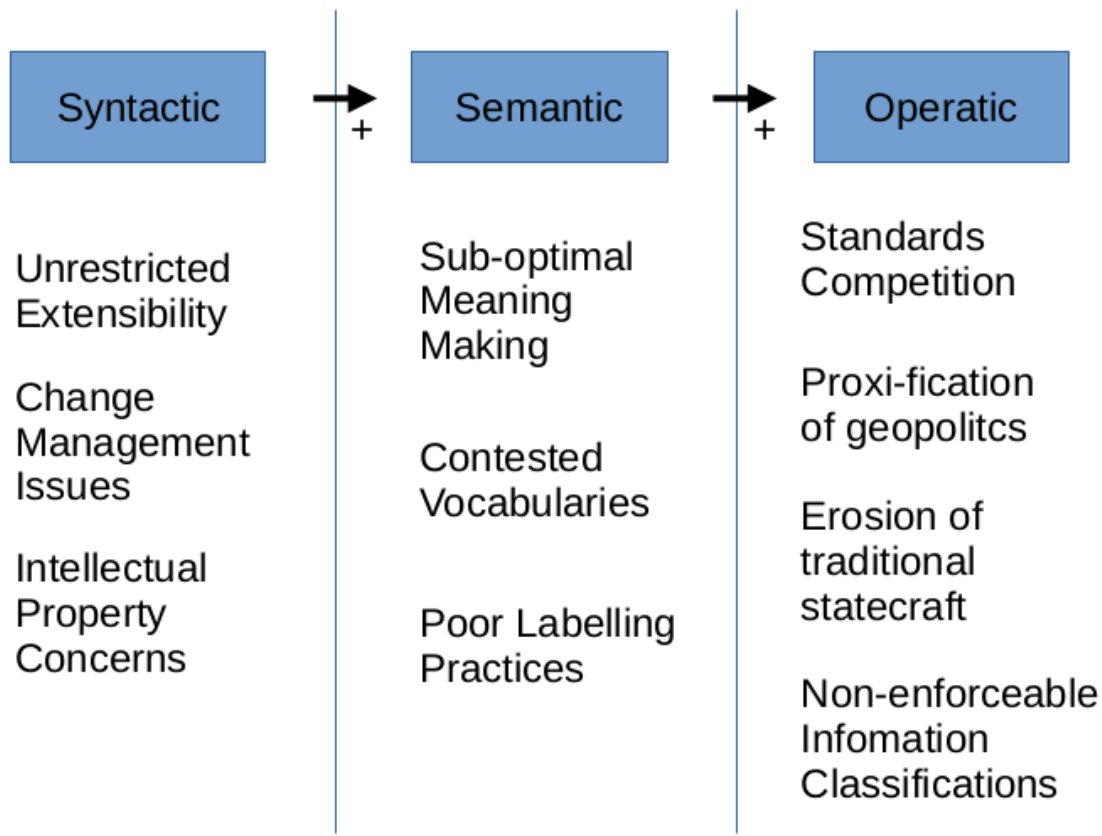

*Figure 4: Summarising the challanges of cyberspace awareness as encountered through the lens of MSA*

Interestingly, these three tie together decisions in organisational data and information policies to their implications in the wider strategic and international politics for nation-states. Moreover, CTI comes with certain preferences and constraints predefined by the underlying language standards and information models. This underlying nervous system for cyberspace awareness is not as agile as it should be, and may require a significant amount of customisation for emergent threats like MSA, which we've indicated for all the three dimensions of cyberspace awareness.

For policymakers, it is therefore essential to take stock of each dimension of cyberspace awareness while adopting a particular CTI mechanism, as these standardised information sharing methods may also carry along their own implications in international politics and governance structures. Further, sharing the CTI of MSA applications in the existing paradigm of cyberspace awareness can also have consequences concerning the privateness and politics of information.

While the sharing of cyber threat information has become an industry in itself, a corresponding mechanism for states to vet this 'knowledge economy' is not only non-existent but also necessary in order to navigate the integration of social automation based threats into CTI.

*****

# APPENDIX A

## Basic Schema for a Situational Event Describing MSA [written in TypeQL]

# APPENDIX B

List of members of OASIS Technical Committee which formulated STIX2.1, with their respective affiliations:

| Name | Organisation | Org Country | Sector |
|---|---|---|---|
| Kai Li | 360 Enterprise Security Group | China | Private Sector |
| Shu Li | 360 Enterprise Security Group | China | Private Sector |
| Qian Yin | 360 Enterprise Security Group | China | Private Sector |
| Xinhua Zheng | 360 Enterprise Security Group | China | Private Sector |
| Robert Coderre | Accenture | Ireland | Private Sector |
| Kyle Maxwell | Accenture | Ireland | Private Sector |
| David Crawford | Aetna | US | Private Sector |
| Marcos Orallo | Airbus Group SAS | Europe | Private Sector |
| Roman Fiedler | AIT Austrian Institute of Technology | Austria | Academia |
| Florian Skopik | AIT Austrian Institute of Technology | Austria | Academia |
| Ryan Clough | Anomali | US | Private Sector |
| Nicholas Hayden | Anomali | US | Private Sector |
| Wei Huang | Anomali | US | Private Sector |
| Russell Matbouli | Anomali | US | Private Sector |
| Angela Nichols | Anomali | US | Private Sector |
| Hugh Njemanze | Anomali | US | Private Sector |
| Katie Pelusi | Anomali | US | Private Sector |
| Patrick Maroney | AT&T | US | Private Sector |
| Dean Thompson | Australia and New Zealand Banking Group (ANZ Bank) | Australia | Private Sector |
| Radu Marian | Bank of America | US | Private Sector |
| Sounil Yu | Bank of America | US | Private Sector |
| Vicky Laurens | Bank of Montreal | Canada | Private Sector |
| Bret Jordan | Broadcom | US | Private Sector |
| Trey Darley | CCB/CERT.be | Belgium | Government |
| Alexandre Dulaunoy | CIRCL | Luxembourg | Government |
| Andras Iklody | CIRCL | Luxembourg | Government |
| Christian Studer | CIRCL | Luxembourg | Government |
| RaphaÎl Vinot | CIRCL | Luxembourg | Government |
| Syam Appala | Cisco Systems | US | Private Sector |
| Ted Bedwell | Cisco Systems | US | Private Sector |
| Pavan Reddy | Cisco Systems | US | Private Sector |
| Omar Santos | Cisco Systems | US | Private Sector |
| Sam Taghavi Zargar | Cisco Systems | US | Private Sector |
| Jyoti Verma | Cisco Systems | US | Private Sector |
| Jart Armin | Cyber Threat Intelligence Network, Inc. (CTIN) | US | Private Sector |
| Doug DePeppe | Cyber Threat Intelligence Network, Inc. (CTIN) | US | Private Sector |
| Jane Ginn | Cyber Threat Intelligence Network, Inc. (CTIN) | US | Private Sector |
| Ben Ottoman | Cyber Threat Intelligence Network, Inc. (CTIN) | US | Private Sector |
| David Powell | Cyber Threat Intelligence Network, Inc. | US | Private Sector |

| | (CTIN) | | |
|---|---|---|---|
| Andreas Sfakianakis | Cyber Threat Intelligence Network, Inc. (CTIN) | US | Private Sector |
| Anuj Goel | Cyware Labs | US | Private Sector |
| Avkash Kathiriya | Cyware Labs | US | Private Sector |
| Jaeden Hampton | DarkLight, Inc. | US | Private Sector |
| Ryan Hohimer | DarkLight, Inc. | US | Private Sector |
| Ryan Joyce | DarkLight, Inc. | US | Private Sector |
| Shawn Riley | DarkLight, Inc. | US | Private Sector |
| Ian Roberts | DarkLight, Inc. | US | Private Sector |
| Andrew Byrne | Dell | US | Private Sector |
| Jeff Odom | Dell | US | Private Sector |
| Sreejith Padmajadevi | Dell | US | Private Sector |
| Ravi Sharda | Dell | US | Private Sector |
| Will Urbanski | Dell | US | Private Sector |
| David Ailshire | DHS Office of Cybersecurity and Communications (CS&C) | US | Government |
| Steven Fox | DHS Office of Cybersecurity and Communications (CS&C) | US | Government |
| Taneika Hill | DHS Office of Cybersecurity and Communications (CS&C) | US | Government |
| Evette Maynard-Noel | DHS Office of Cybersecurity and Communications (CS&C) | US | Government |
| Jackie Eun Park | DHS Office of Cybersecurity and Communications (CS&C) | US | Government |
| Sean Sobieraj | DHS Office of Cybersecurity and Communications (CS&C) | US | Government |
| Marlon Taylor | DHS Office of Cybersecurity and Communications (CS&C) | US | Government |
| Preston Werntz | DHS Office of Cybersecurity and Communications (CS&C) | US | Government |
| Jˆrg Abraham | EclecticIQ | Netherlands | Private Sector |
| wouter bolsterlee | EclecticIQ | Netherlands | Private Sector |
| Adam Bradbury | EclecticIQ | Netherlands | Private Sector |
| Marko Dragoljevic | EclecticIQ | Netherlands | Private Sector |
| Oliver Gheorghe | EclecticIQ | Netherlands | Private Sector |
| Joep Gommers | EclecticIQ | Netherlands | Private Sector |
| Caitlin Huey | EclecticIQ | Netherlands | Private Sector |
| Christopher O'Brien | EclecticIQ | Netherlands | Private Sector |
| Sergey Polzunov | EclecticIQ | Netherlands | Private Sector |
| Rutger Prins | EclecticIQ | Netherlands | Private Sector |
| Aukjan van Belkum | EclecticIQ | Netherlands | Private Sector |
| Raymon van der Velde | EclecticIQ | Netherlands | Private Sector |
| Tom Vaughan | EclecticIQ | Netherlands | Private Sector |
| Joseph Woodruff | EclecticIQ | Netherlands | Private Sector |
| Ben Sooter | Electric Power Research Institute (EPRI) | US | Non-Profit |
| Chris Ricard | Financial Services Information Sharing and Analysis Center (FS-ISAC) | US | Non-Profit |
| Sean Barnum | FireEye, Inc. | US | Private Sector |
| Phillip Boles | FireEye, Inc. | US | Private Sector |
| Prasad Gaikwad | FireEye, Inc. | US | Private Sector |
| Haripriya Gajendran | FireEye, Inc. | US | Private Sector |

| Will Green | FireEye, Inc. | US | Private Sector |
|---|---|---|---|
| Rajeev Jha | FireEye, Inc. | US | Private Sector |
| Gary Katz | FireEye, Inc. | US | Private Sector |
| Anuj Kumar | FireEye, Inc. | US | Private Sector |
| James Meck | FireEye, Inc. | US | Private Sector |
| Shyamal Pandya | FireEye, Inc. | US | Private Sector |
| Paul Patrick | FireEye, Inc. | US | Private Sector |
| Remko Weterings | FireEye, Inc. | US | Private Sector |
| Tim Jones | ForeScout | US | Private Sector |
| Ryusuke Masuoka | Fujitsu Limited | Japan | Private Sector |
| Daisuke Murabayashi | Fujitsu Limited | Japan | Private Sector |
| Derek Northrope | Fujitsu Limited | Japan | Private Sector |
| Toshitaka Satomi | Fujitsu Limited | Japan | Private Sector |
| Koji Yamada | Fujitsu Limited | Japan | Private Sector |
| Kunihiko Yoshimura | Fujitsu Limited | Japan | Private Sector |
| Robert van Engelen | Genivia | US | Private Sector |
| Eric Burger | Georgetown University | US | Academia |
| Allison Miller | Google Inc. | US | Private Sector |
| Mark Risher | Google Inc. | US | Private Sector |
| Yoshihide Kawada | Hitachi, Ltd. | Japan | Private Sector |
| Jun Nakanishi | Hitachi, Ltd. | Japan | Private Sector |
| Kazuo Noguchi | Hitachi, Ltd. | Japan | Private Sector |
| Akihito Sawada | Hitachi, Ltd. | Japan | Private Sector |
| Yutaka Takami | Hitachi, Ltd. | Japan | Private Sector |
| Masato Terada | Hitachi, Ltd. | Japan | Private Sector |
| Adrian Bishop | Huntsman Security | Australia | Private Sector |
| Eldan Ben-Haim | IBM | US | Private Sector |
| Allen Hadden | IBM | US | Private Sector |
| Sandra Hernandez | IBM | US | Private Sector |
| Jason Keirstead | IBM | US | Private Sector |
| Chenta Lee | IBM | US | Private Sector |
| John Morris | IBM | US | Private Sector |
| Devesh Parekh | IBM | US | Private Sector |
| Emily Ratliff | IBM | US | Private Sector |
| Nick Rossmann | IBM | US | Private Sector |
| Laura Rusu | IBM | US | Private Sector |
| Ron Williams | IBM | US | Private Sector |
| Paul Martini | iboss, Inc. | US | Private Sector |
| Vasileios Mavroeidis | IFI | Norway | Academia |
| Kamer Vishi | IFI | Norway | Academia |
| Joerg Eschweiler | Individual | Germany | NA |
| Elysa Jones | Individual | US | NA |
| Terry MacDonald | Individual | Australia | NA |
| Tim Casey | Intel Corporation | US | Private Sector |
| Julie Modlin | Johns Hopkins University Applied Physics Laboratory | US | Academia |
| Mark Moss | Johns Hopkins University Applied Physics Laboratory | US | Academia |
| Mark Munoz | Johns Hopkins University Applied Physics Laboratory | US | Academia |
| Nathan Reller | Johns Hopkins University Applied Physics | US | Academia |
| | Laboratory | | |
| Pamela Smith | Johns Hopkins University Applied Physics Laboratory | US | Academia |
| Vivek Jain | JPMorgan Chase Bank, N.A. | US | Private Sector |
| Subodh Kumar | JPMorgan Chase Bank, N.A. | US | Private Sector |
| David Laurance | JPMorgan Chase Bank, N.A. | US | Private Sector |
| Russell Culpepper | Kaiser Permanente | US | Private Sector |
| Beth Pumo | Kaiser Permanente | US | Private Sector |
| Michael Slavick | Kaiser Permanente | US | Private Sector |
| Daniel Ben-Chitrit | Looking Glass | US | Private Sector |
| Wesley Brown | Looking Glass | US | Private Sector |
| Dennis Hostetler | Looking Glass | US | Private Sector |
| Himanshu Kesar | Looking Glass | US | Private Sector |
| Matt Pladna | Looking Glass | US | Private Sector |
| Vlad Serban | Looking Glass | US | Private Sector |
| Allan Thomson | Looking Glass | US | Private Sector |
| Chris Wood | Looking Glass | US | Private Sector |
| Kent Landfield | McAfee | US | Private Sector |
| Jonathan Baker | Mitre Corporation | US | Non-Profit |
| Desiree Beck | Mitre Corporation | US | Non-Profit |
| Michael Chisholm | Mitre Corporation | US | Non-Profit |
| Sam Cornwell | Mitre Corporation | US | Non-Profit |
| Sarah Kelley | Mitre Corporation | US | Non-Profit |
| Ivan Kirillov | Mitre Corporation | US | Non-Profit |
| Michael Kouremetis | Mitre Corporation | US | Non-Profit |
| Chris Lenk | Mitre Corporation | US | Non-Profit |
| Nicole Parrish | Mitre Corporation | US | Non-Profit |
| Richard Piazza | Mitre Corporation | US | Non-Profit |
| Larry Rodrigues | Mitre Corporation | US | Non-Profit |
| Jon Salwen | Mitre Corporation | US | Non-Profit |
| Charles Schmidt | Mitre Corporation | US | Non-Profit |
| Richard Struse | Mitre Corporation | US | Non-Profit |
| Alex Tweed | Mitre Corporation | US | Non-Profit |
| Emmanuelle Vargas-Gonzalez | Mitre Corporation | US | Non-Profit |
| John Wunder | Mitre Corporation | US | Non-Profit |
| James Cabral | MTG Management Consultants, LLC. | US | Private Sector |
| Scott Algeier | National Council of ISACs (NCI) | US | Non-Profit |
| Denise Anderson | National Council of ISACs (NCI) | US | Non-Profit |
| Josh Poster | National Council of ISACs (NCI) | US | Non-Profit |
| Mike Boyle | National Security Agency | US | Government |
| Jessica Fitzgerald-McKay | National Security Agency | US | Government |
| David Kemp | National Security Agency | US | Government |
| Shaun McCullough | National Security Agency | US | Government |
| Jason Romano | National Security Agency | US | Government |
| John Anderson | NC4 | US | Private Sector |
| Michael Butt | NC4 | US | Private Sector |
| Mark Davidson | NC4 | US | Private Sector |
| Daniel Dye | NC4 | US | Private Sector |
| Michael Pepin | NC4 | US | Private Sector |

| Natalie Suarez | NC4 | US | Private Sector |
|---|---|---|---|
| Benjamin Yates | NC4 | US | Private Sector |
| Sarah Brown | NCI Agency | NATO | Government |
| Oscar Serrano | NCI Agency | NATO | Government |
| Daichi Hasumi | NEC Corporation | Japan | Private Sector |
| Takahiro Kakumaru | NEC Corporation | Japan | Private Sector |
| Lauri Korts-Parn | NEC Corporation | Japan | Private Sector |
| Kelly Cullinane | New Context Services, Inc. | US | Private Sector |
| John-Mark Gurney | New Context Services, Inc. | US | Private Sector |
| Christian Hunt | New Context Services, Inc. | US | Private Sector |
| Danny Purcell | New Context Services, Inc. | US | Private Sector |
| Daniel Riedel | New Context Services, Inc. | US | Private Sector |
| Andrew Storms | New Context Services, Inc. | US | Private Sector |
| Drew Varner | NineFX, Inc. | US | Private Sector |
| Stephen Banghart | NIST | US | Government |
| David Darnell | North American Energy Standards Board | US | Non-Profit |
| James Crossland | Northrop Grumman | US | Private Sector |
| Robert Van Dyk | Northrop Grumman | US | Private Sector |
| Cheolho Lee | NSRI | US | Academia |
| Cory Casanave | Object Management Group | US | Non-Profit |
| Joel Myhre | Pacific Disaster Center | US | Government |
| Vishaal Hariprasad | Palo Alto Networks | US | Private Sector |
| Brad Bohen | Perch | US | Private Sector |
| Aharon Chernin | Perch | US | Private Sector |
| Zach Kanzler | Perch | US | Private Sector |
| Michael Lane | Perch | US | Private Sector |
| Michael Riggs | Perch | US | Private Sector |
| Sean O'Brien | Purism SPC | US | Private Sector |
| John Tolbert | Queralt Inc. | US | Private Sector |
| Forrest Hare | Science Application International | US | Private Sector |
| Duncan Sparrell | sFractal Consulting LLC | US | Private Sector |
| Thomas Schreck | Siemens AG | Germany | Private Sector |
| Adam Wyner | Swansea University | UK | Academia |
| Bret Jordan | Symantec Corp. | US | Private Sector |
| Robert Keith | Symantec Corp. | US | Private Sector |
| Curtis Kostrosky | Symantec Corp. | US | Private Sector |
| Chris Larsen | Symantec Corp. | US | Private Sector |
| Michael Mauch | Symantec Corp. | US | Private Sector |
| Aubrey Merchant | Symantec Corp. | US | Private Sector |
| Efrain Ortiz | Symantec Corp. | US | Private Sector |
| Mingliang Pei | Symantec Corp. | US | Private Sector |
| Kenneth Schneider | Symantec Corp. | US | Private Sector |
| Arnaud Taddei | Symantec Corp. | US | Private Sector |
| Brian Witten | Symantec Corp. | US | Private Sector |
| Greg Reaume | TELUS | Canada | Private Sector |
| Alan Steer | TELUS | Canada | Private Sector |
| Crystal Hayes | The Boeing Company | US | Private Sector |
| Andrew Gidwani | ThreatConnect, Inc. | US | Private Sector |
| Cole Iliff | ThreatConnect, Inc. | US | Private Sector |
| Andrew Pendergast | ThreatConnect, Inc. | US | Private Sector |
| +Jason Spies | ThreatConnect, Inc. | US | Private Sector |

| Ryan Trost | ThreatQuotient, Inc. | US | Private Sector |
|---|---|---|---|
| David Girard | Trend Micro | Japan | Private Sector |
| Brandon Niemczyk | Trend Micro | Japan | Private Sector |
| Eric Shulze | Trend Micro | Japan | Private Sector |
| Patrick Coughlin | TruSTAR Technology | US | Private Sector |
| Chris Roblee | TruSTAR Technology | US | Private Sector |
| ADHAM ALBAKRI | University of Kent | UK | Academia |
| Jeffrey Mates | US Department of Defense (DoD) | US | Government |
| Evette Maynard-Noel | US Department of Homeland Security | US | Government |
| Lee Chieffalo | Viasat | US | Private Sector |
| Wilson Figueroa | Viasat | US | Private Sector |
| Andrew May | Viasat | US | Private Sector |
| Ales Cernivec | XLAB | Slovenia | Private Sector |
| Anthony Rutkowski | Yanna Technologies LLC | US | Private Sector |